\documentclass[12pt,preprint]{aastex}

\usepackage{natbib}

\newcount \note  \note=0

\def\rmit#1{{\it #1}}              %
\def\specchar#1{{\sc #1}}
\def\FeI{\mbox{Fe\,\specchar{i}}}

\def\BeII{\mbox{Be\,\specchar{ii}}}

\def\ie{\rmit{i.e.}}
\def\eg{\rmit{e.g.}}

\shorttitle{Solar abundance corrections from 3D MHD simulations}
\shortauthors{Fabbian et al.}

\begin{document}

\title{Solar abundance corrections \\ derived through 3D magnetoconvection simulations }

\author{D. Fabbian\altaffilmark{*}, E. Khomenko\altaffilmark{*}, F. Moreno-Insertis\altaffilmark{*}}
\affil{Instituto de Astrofisica de Canarias (IAC), E-38200 La Laguna, Tenerife, Spain}
\email{[damian;khomenko;fmi]@iac.es}

\and

\author{\rm{\AA}. Nordlund}
\affil{Niels Bohr Institute, University of Copenhagen,
Blegdamsvej 17, DK-2100 Copenhagen, Denmark} \email{aake@nbi.dk}

\altaffiltext{*}{Also affiliated with: Departamento de
Astrofisica, Universidad de La Laguna (ULL), E-38205 La
Laguna, Tenerife, Spain}

\begin{abstract}
  We explore the effect of the magnetic field when using realistic
  three-dimensional convection experiments to determine solar element
  abundances.  By carrying out magnetoconvection simulations with a
  radiation-hydro code (the Copenhagen stagger code) and through a-posteriori
  spectral synthesis of three \FeI\ lines, we obtain evidence that
  moderate amounts of mean magnetic flux cause a noticeable 
  change in the derived equivalent widths compared with those for a
  non-magnetic case. The corresponding Fe abundance correction for a mean
  flux density of $200$ G reaches up to $\sim 0.1$ dex. These
  results are based on space- and time-averaged line profiles over a time
  span of $2.5$ solar hours in the statistically stationary regime
  of the convection. The main factors causing the change in equivalent
  widths, namely the Zeeman broadening and the modification of the temperature
  stratification, act in different amounts and, for the iron lines
    considered here, in opposite directions; yet, the resulting
  $|\Delta\log\epsilon_{\odot}(Fe)|$ coincides within a factor two
  in all of them, even though the sign of the total abundance correction is different
  for the visible and infrared lines.  We conclude that magnetic
  effects should be taken into account when discussing precise values of the
  solar and stellar abundances and that an extended study is warranted.
\end{abstract}


\section{Introduction}\label{sec:intro}

The advent of increasingly realistic three-dimensional convection
simulations based on radiation-hydrodynamics codes
\citep{Dravinsetal1981,Nordlund1985,Freytagetal2002,
Vogleretal2005} has led to the conclusion that horizontal
inhomogeneities and departures from Local Thermodynamic
Equilibrium (LTE) are crucial for an accurate abundance
determination \citep[see the review by][]{Asp2005}.
Three-dimensional hydrodynamic (HD) models have allowed to remove many
uncertainties in abundance determination, in particular those
coming from the use of micro- and macroturbulence velocity
parameters \citep{Aspetal2000a}. Yet, several sources of uncertainty remain, resulting,
e.g., from possible blending of lines, errors in continuum
measurements, equivalent width determination
\citep{HolwKockBard1995, BlackSmithLyn1995, KostShchuRutt1996,
Ayres2008a, Ayres2008b, Caffetal2008} or calculation of collisions
with H atoms \citep{Fabbianetal2006, Fabbianetal2009}.
Another potentially important source of uncertainty stems from the
presence of significant amounts of magnetic flux in the photosphere of
the quiet Sun \citep{Stenflo1982, SanchezAlmeida_etal_2003,
Shchukina_TrujilloBueno2003, TrujilloBueno+etal2004} and its possible
influence on the average width of the line profiles
\citep{Stenflo1977}. The purpose of this Letter is to use 3D
magnetoconvection (MHD) models to estimate the influence of the
magnetic field on the spectral profiles of a number of \FeI\
lines, and to translate the difference between the equivalent widths in
MHD and HD models to the corresponding Fe abundance corrections.

The impact of magnetic fields on chemical abundance determinations
has so far been widely neglected.
It was implicitly assumed that the
magnetic field should not play a major role in changing the shapes of
spatially-averaged spectral lines, as these averages are weighted
towards granules \citep{NordStAsp2009}, while magnetic
concentrations are thought to reside mainly in intergranular lanes
\citep{Khomenko+etal2003}.
Still, \citet{Borrero2008} found that the neglect of magnetic
broadening can lead to a systematic error of up to $0.1$ dex in the abundance
determination of iron.
The author used 1D spectral synthesis with uniform magnetic field in a plane-parallel atmosphere,
therefore inevitably neglecting the effect of the field on the
temperature stratification.
However, the latter effect, although indirect, can be important \citep[see,
e.g.][]{Socas-Navarro&Norton2007} and must be considered alongside the
direct effect through Zeeman broadening \citep[see, e.g.][]{Stenflo1977}.

Iron is one of the crucial chemical elements for a variety of
reasons like: (a) it is widely used to define a metallicity scale,
hence as a proxy for the overall metal content;
(b) the effect of the magnetic field on the line profiles seem to be
more pronounced for this element than for,
\eg\ Si, C and O \citep{Borrero2008}; (c) the large amount of iron
lines in the solar spectrum allows to consider features with high
Land\'e factor. It is therefore advisable to start with iron
when studying the magnetic effects on abundance determinations from MHD models.
In this Letter, we present results based on a series of
three-dimensional magneto-convection simulations. 
Our aim is to explore, understand and
constrain the possible impact on the solar abundance determination
of self-consistent, simultaneous 3D and magnetic effects.
These effects are studied via comparison of the equivalent widths
for the magnetic cases with those obtained in HD conditions. To derive the
necessary abundance correction, we match the MHD equivalent width with that
obtained on the basis of HD runs with changed abundance. Our
analysis focuses on quiet Sun conditions.
We conclude that magnetic effects, in particular via
modification of the average temperature stratification, should be taken into
account when considering the solar abundance problem.

\section{Simulations and spectral synthesis}

\begin{figure*}
\centering
%
%\epsscale{2.00}
\plotone{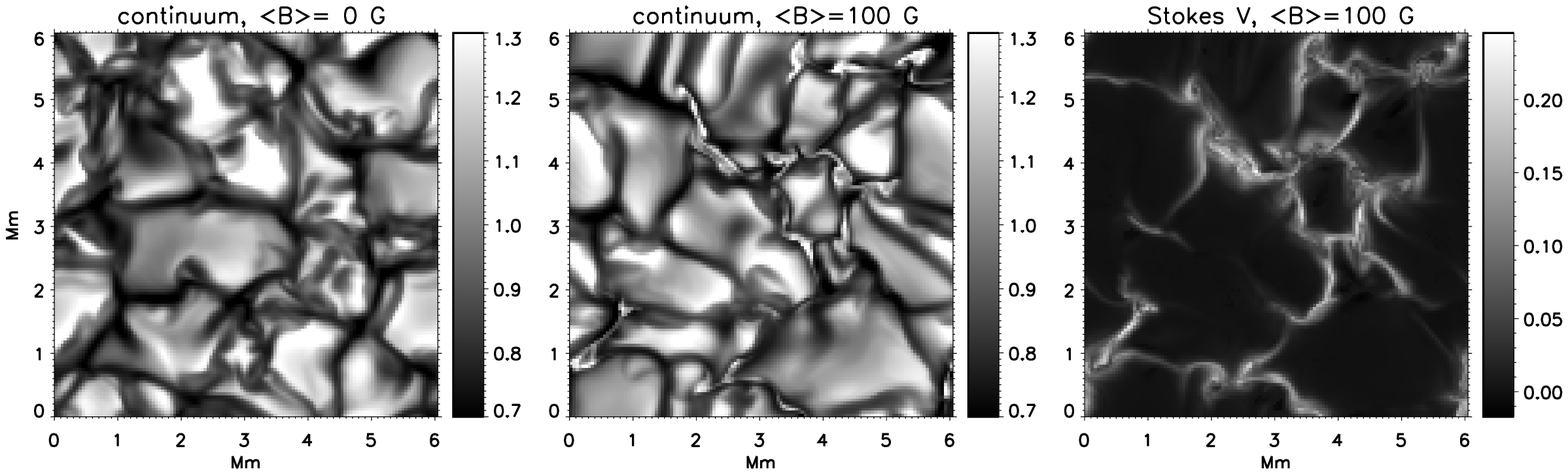}
\caption{Emerging continuum intensity at 608 nm in the HD
simulation run (left), and in the MHD run with ``100 G''
setup (center).
Note the appearance of intergranular bright points
after introducing magnetic flux.
The right panel shows the amplitude of Stokes V, defined as
max($|\textrm{V}|$) in units of the continuum intensity, sign included.
} \label{fig:contin}
\end{figure*}

The results in this paper are based on simulations we carried out
using the Copenhagen stagger-code
\citep{nordlund+galsgaard95,Carlssonetal2004,Stein&Nord2006,DePontieuetal2006},
a high-order finite-difference Cartesian MHD code using
hyper-diffusivities.  All calculations reported here are for a domain
of $6 \times 3 \times 6$ Mm$^3$, with a grid of $252 \times 126 \times
252$ points and uniform spacing in the horizontal directions. The
vertical direction is non-uniform and reaches a maximum resolution of
$\sim$15 km in the photosphere. This grid should provide sufficient
resolution for the purpose of abundance studies \citep{Aspetal2000b}.

To initiate the MHD runs, a vertical uniform magnetic field of
strength $B_0$ was introduced into an already evolved HD snapshot. An
HD run and three different MHD cases (with $B_0 = 50, 100, 125$ G)
were evolved for up to several solar hours, with snapshots taken every
30 seconds. When the implanted magnetic field is sufficiently strong,
as the simulation evolves the continuum intensity starts to show fine
bright threads/points and fragments (Fig.~\ref{fig:contin}), as in
high-resolution solar observations
\citep[e.g.][]{SanchezAlmeida_etal_2004, SanchezAlmeida_etal_2010,
shelyag_etal_2004}.

To simplify the task, the spectral synthesis was performed in LTE.  We
considered the three lines of \FeI\ at $\lambda = 608.27$, $624.07$
and $1564.85$ nm, which have EPL $= 2.223 $, $2.223$ and $5.43$ eV;
log gf = $-3.572, -3.390, -0.670$ and $g_{\rm eff}=2.00, 0.99 $ and
$2.98$, respectively.  We used the LILIA code
\citep{Socas-Navarro2001}, and set the resolution to 1.28 pm over 200
wavelength points.

The synthetic spectra were computed using no enhancement factor to the
Van der Waals broadening formula in the spectral synthesis.
To obtain the final synthetic profiles, emergent
spectra were calculated for each vertical column in the grid. Averages
were then calculated for each wavelength over the horizontal
directions and in time, using $95$ solar minutes of the stationary
stage of the simulations.

\section{Results}\label{sect:results}

\subsection{Goodness of the simulations}

\begin{figure}
%
%
%\centering
%
%\epsscale{1.05}
\plotone{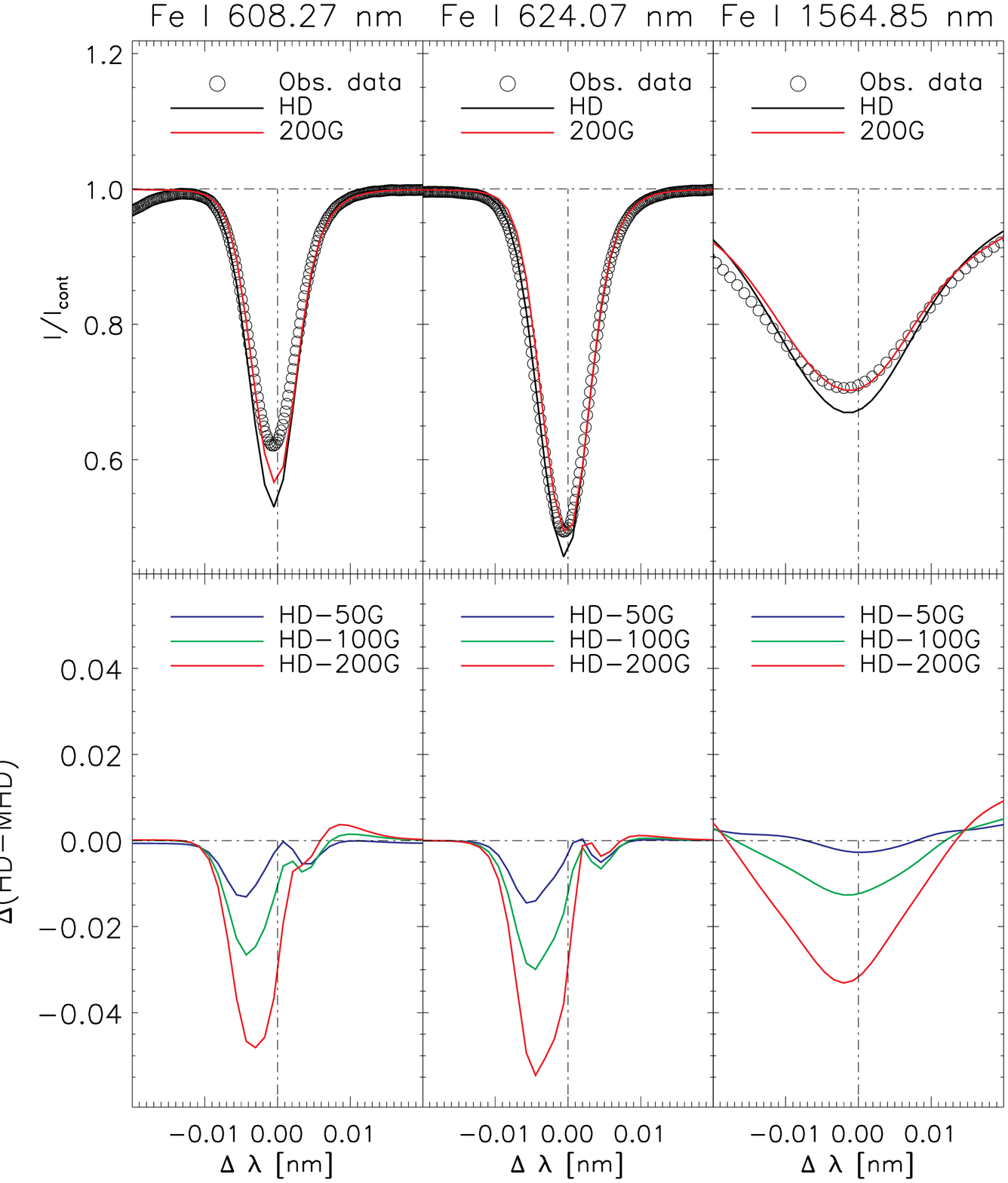}
\caption{{\it Top panels}: normalized theoretical profiles after spatial- and temporal-average
for the \FeI\ $608.27$ nm (left), $624.07$ nm (center) and (right) $1564.85$ nm spectral lines,
as derived from the synthesis of relevant spectral regions from the HD and MHD simulations.
For the sake of clarity, MHD results are shown only for the simulation with highest magnetic flux.
Spectral observational data from the Kitt Peak FTS Spectral Atlas
are given for visual comparison (empty circles).
{\it Bottom panels}: intensity difference between average normalized synthetic line profiles in HD and MHD.}
\label{fig_MHD}
\end{figure}

In Fig.~\ref{fig_MHD}, we show (solid lines) the spatially- and
  temporally-averaged synthetic line profiles from our HD and $200$ G experiments (top
  panels), and the differences in intensity between the spectral profiles derived in the MHD cases
  and those obtained for the HD case
  (bottom panels).
  Each of the profiles shown was normalized {\it to its
  own continuum level}.

Observational data taken from the FTS spectral atlas
\citep{Braultetal1987} are also shown in the figure for visual
comparison (top panels, empty circles). The fit of the line profiles
is satisfactory, with the MHD models performing better than the HD
one. In any case, since the objective is a {\it differential} analysis
between MHD and HD, the scope of the work presented here does not
require obtaining extremely high accuracy in the fitting of observed
line profiles. Remaining discrepancies in the fitting, especially
in the line cores, can be explained by the higher temperatures in the
relevant layers of our MHD models, which will lead to stronger neutral
iron overionisation and are thus likely to induce larger departures
from the LTE assumption adopted here.

More importantly with regards to our aim here, we verified that our
equivalent width calculation converged with regards to the number of
sampled spectral wavelengths. Moreover, our equivalent-width estimates
are in fair agreement with the observational values found in the
literature \citep{GurtovenkoKostik, BlackLynSmith1995,
HolwKockBard1995, KostShchuRutt1996}.

A further check of the goodness of our simulations is possible: we
obtain an excellent match between our continuum intensity values and
those expected from the literature \citep[see figure 2
of][]{TruShchu2009}.  For example, in units of erg/(cm$^2$ s ster
\AA), we get Stokes I values of $\sim 3.13 \cdot 10^{6}$ (continuum
around the \BeII\ $313.04$ and $313.11$ nm lines), $\sim 4.88 \cdot
10^{6}$ (continuum around the \FeI\ 401.02 nm line), $\sim 2.91 \cdot
10^{6}$ (around the \FeI\ $630.15$ and $630.25$ nm lines) and $\sim
2.43 \cdot 10^{6}$ (continuum around the \FeI\ 700.06 nm line). The
latter two values match almost perfectly with the observational data
of \citet{AyrPlyKell(2006)}. On the other hand, our UV values, while
still agreeing to a good level, may be slightly too high due to our
neglect of line haze on continuum at those wavelengths.

We have also used Stokes V polarization signature results
(Fig.~\ref{fig:contin}, right panel) to derive Stokes V
amplitudes. Depending on the wavelength considered, we find Stokes V
average amplitudes (measured in units of Stokes I continuum intensity)
of the order of $0.01-0.03 \%$ ("100 G" case). This corresponds, when
calibrating our Stokes V signals using the weak-field approximation,
\citep[e.g][]{MarGonBellRub2009}, to an average unsigned vertical
magnetic flux of the order of $10-50$ G. These numbers are typical for
solar internetwork and network regions, as deduced when considering
the effect of limited spatial resolution of observations. This result
confirms that our simulations are representative of solar quiet
regions, a description covering 90 percent of the solar surface.

Regarding limb-darkening, recent work by \citet{PerAspKis2009a} shows
that related uncertainties in the simulations play only a minimal role for
solar abundances. Our differential approach ensures that the small impact of
any discrepancy found against limb darkening measurements can be neglected here.

\subsection{Thermal versus magnetic effects}\label{subsect:effects}

\begin{figure}
   \plotone{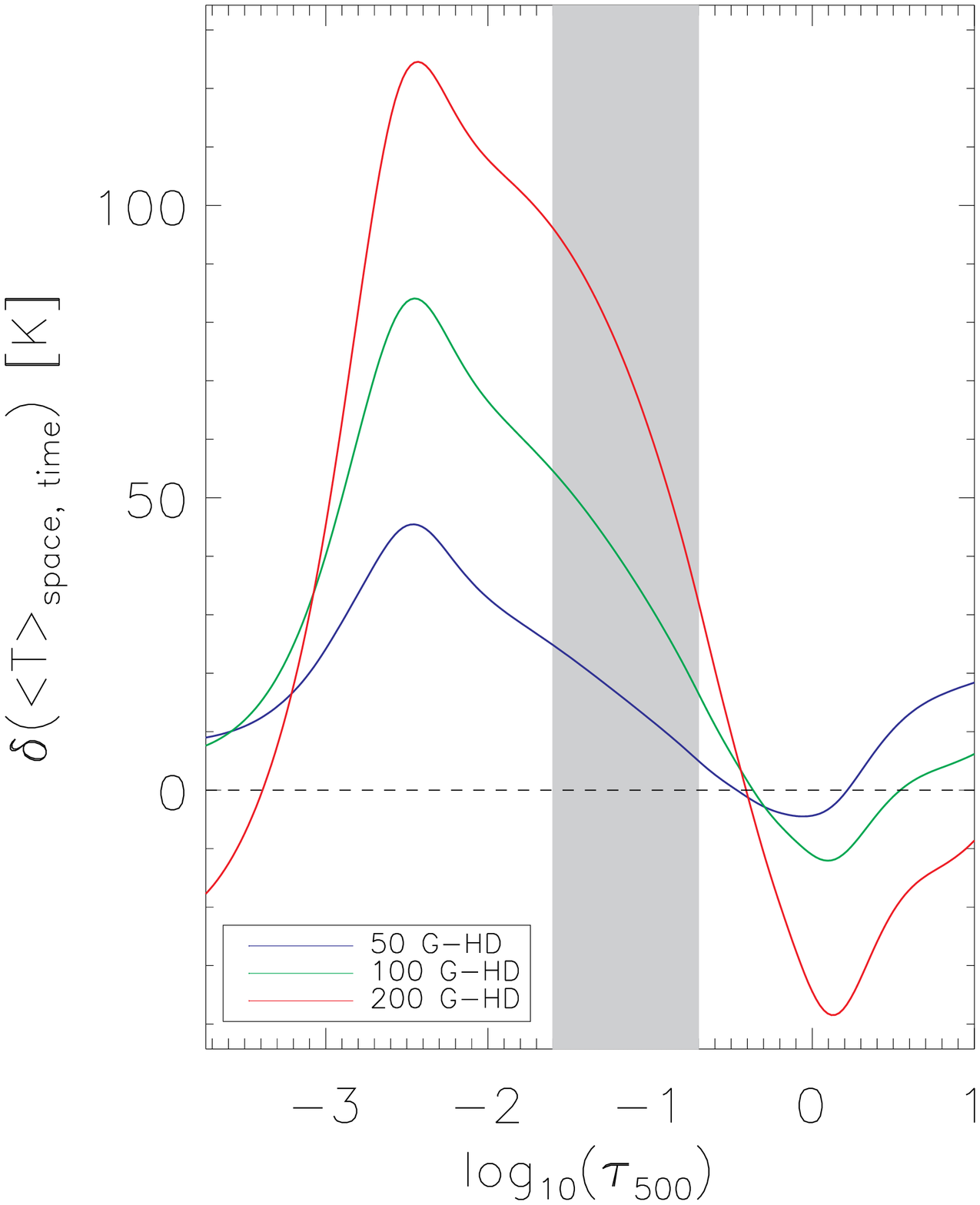}
   \caption{Differences with the HD case in the temperature
     stratification $\langle T \rangle$ (with averaging taken at fixed optical depth, 
     and over time), for $B_0=50$ G (blue line), $100$ G (green line), and
     $200$ G (red line) cases, plotted against a $log_{10}(\tau_{500})$ scale.
     The approximate photospheric region of formation for the lines studied is shown 
     with a gray vertical band.
     \label{fig_stratif}}
\end{figure}

Figure \ref{fig_stratif} compares the average temperature
stratification of the different magnetic runs versus optical depth:
the averaging was done on a fixed $\log(\tau_{500})$ scale, and
over all snapshots. As seen in the figure, in the relevant layers of
formation of the absorption features we consider, $\langle T \rangle$
increases with increasing magnetic flux in the simulations, a result
expected from radiation transfer considerations in strong solar
magnetic elements \citet{Spruit1976,SchSol1988}. 

The formation heights of the relevant lines, for the HD case,
are: -0.276 Mm for \FeI\ $624.07$ nm and -0.229 Mm for \FeI\ $608.27$
nm \citep{GurtovenkoKostik}, and -0.135 Mm for \FeI\ $1564.85$ nm
\citet{ShchuTrujilloBueno2001}, which correspond to an
optical depth of between $\sim -1.6$ and $\sim -0.7$. As seen in the
figure, at these optical depths there is a significant temperature
increase in the MHD cases compared to HD. The maximum temperature 
difference, arises at
$\log(\tau_{500}) \sim -2.5$, reaching up to $125$ K for the run with average vertical magnetic 
flux density of $200$ G.
Together with the direct effect due to the presence of a magnetic field, this has
direct implications for the equivalent width of the average profiles and, thus, for the abundance
corrections.

As seen in Fig.~\ref{fig_MHD}, the line core tends to get weaker in the MHD cases for all
  of the Fe lines considered, due to the increased average temperature
  of the corresponding atmospheric models (Fig.~\ref{fig_stratif}).
  For the two visible lines,
  this indirect influence on the line formation turns out to be the
  dominant effect, with the direct effect of the magnetic field on the
  lines being small. Due to its IR wavelength and large magnetic-sensitivity,
  the $1564$ nm line experiences instead significant direct Zeeman broadening,
  which, in terms of resulting equivalent width, goes in the opposite
  direction of the indirect (core-weakening) effect\footnote{Note that this line is formed
  significantly deeper, where the temperature increase compared to the HD case is less significant.}

\subsection{Differential abundance corrections}

\begin{figure*}
\plotone{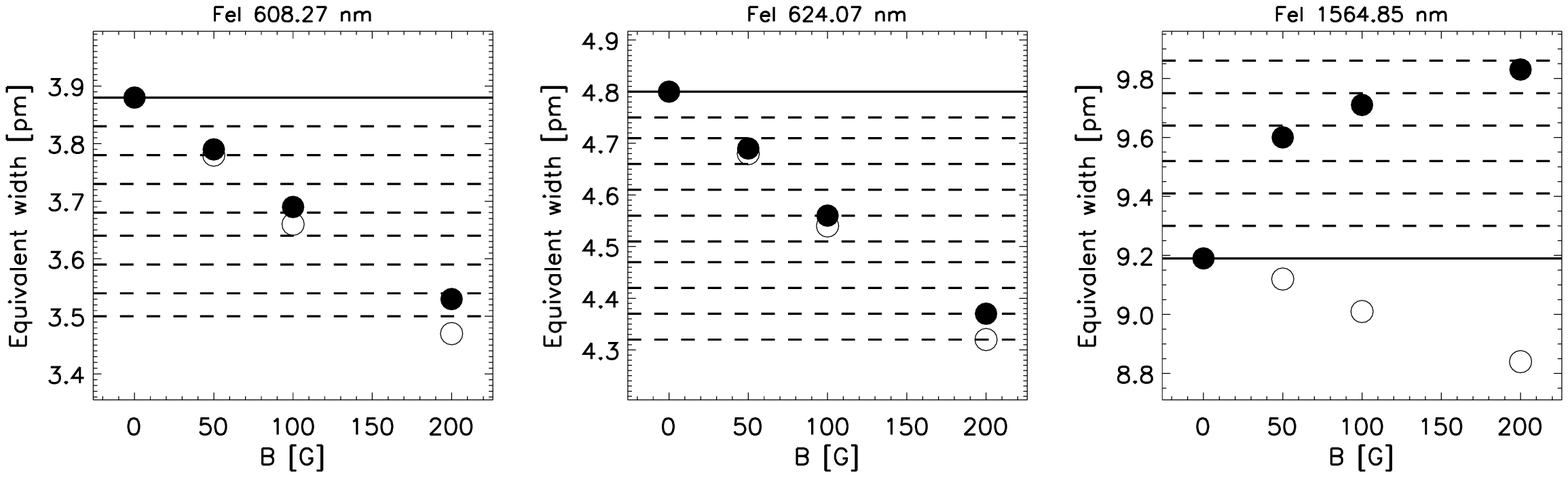} \caption{Filled circles: derived equivalent widths in the HD run and
  the three MHD runs, with separate panels for each spectral line.
Empty circles: equivalent width in the experiments, when doing the spectral
synthesis while setting $B=0$ (leaving the thermodynamics unchanged).
The horizontal lines mark the equivalent
width for the HD case using the standard solar Fe abundance  (solid line),
and for iron abundance changes in steps of $|\Delta\log\epsilon_{\odot}(Fe)| = 0.01$ dex.
\label{fig:ew}}
\end{figure*}

To obtain a differential abundance correction between the HD
and MHD cases, we modify the Fe abundance upward or downward 
of the standard value $\log\epsilon_{\odot}(Fe) = 7.50$
\citep{ShchuTrujilloBueno2001,Aspetal2009} in the
  calculation of the average spectra for the HD series.
  This is done in steps of $0.01$ dex, until the corresponding theoretical equivalent width 
  matches the one derived using 
  the standard iron abundance when computing the average spectra for the MHD case of interest.
  This method allows us to translate the effects of the magnetic
  field into abundance corrections.

The equivalent widths of the three spectral lines calculated with the
standard Fe abundance are presented in Fig.~\ref{fig:ew} (filled circles).
The figure also gives (dashed horizontal 
lines) the equivalent width values for the HD case calculated with the iron
abundance changed in steps of $0.01$ dex.  Since for each case we normalise
the average profiles to the relevant continuum level, changes in equivalent
width are due to actual changes in the shape and depth of the absorption
features.  By comparing the filled circles and the horizontal lines in each
of the three panels of Fig.~\ref{fig:ew}, we see that the maximum abundance
difference is reached in all cases for the runs with the highest field
strength, $B = 200$ G.  The actual numerical values for the maximum
difference are:
$|\Delta\log\epsilon_{\odot}(Fe)| \sim
0.06$ dex for \FeI\ $1564.85$ nm; 
$\sim 0.09$ dex  for \FeI\ $624.07$ nm and slightly above  $\sim 0.07$ dex 
for \FeI\ $608.27$ nm.

The variations of the equivalent width in Fig.~\ref{fig:ew} are caused both
by direct magnetic effects (\ie, via Zeeman broadening) and, indirectly, via
changes in the stratification leading to different amount of absorption in the line core. As
the name implies, Zeeman broadening always causes an increase in the
equivalent width. In contrast, the indirect contribution resulting from
  the increased average temperatures in the MHD models is a line-weakening
  effect for the three chosen lines. This is to be expected \citep*[see
  discussion in][Chapter 13]{Gray1992} for neutral Fe lines in the solar
  photosphere, where iron is mostly ionized in the relevant layers. One can
estimate the relative importance of the direct and indirect effects by ad-hoc
setting $B_{0}=0$ while keeping all other variables unmodified when
  doing the spectral synthesis for the MHD snapshots.
The open circles in Fig.~\ref{fig:ew} show
the equivalent widths resulting from carrying out this experiment for all three lines. For the visible lines
the indirect effects dominate, which can be expected since they form at
heights in the photosphere where the temperature differences between the MHD and HD
models become more important.
The direct effect for those lines leads to a roughly linear increase of
equivalent width with $B$, which is compatible with the Zeeman
broadening origin of this effect.
For the infrared \FeI\ $1564.85$ nm line, in turn, the direct magnetic
effects become dominant 
to the point that they reverse the sign of the
  abundance correction in that case. The sensitivity of this spectral line
to the magnetic field is in fact very high, given its longer wavelength
  and the fact that its Land\'e factor is g$_L \sim 3$. Also, for
  this infrared \FeI\ line 
  there seems to be a saturation effect, although no asymptotic horizontal
  regime is reached in the range of B used in the present study.

\section{Conclusions}

In the past decade, an intense debate has arisen due to the
conflict created by new solar abundance determinations, see
e.g. \citet{Asp2005}, \citet{Ayres2008b}, \citet{Caffetal2008},
\citet{PerKisAsp2009}, \citet{PerAspKis2009b},
\citet{Caffetal2010}. Despite the fact that the new abundances well match meteoritic
values and improve the agreement with solar-neighbourhood and local
interstellar-medium abundances, the proposed large reduction of the
solar overall metallicity has led to controversies, especially from the
helioseismology community, because it spoils the
previous agreement between interior sound speed profiles derived from
helioseismic measurements and predictions based on the classical
solar model \citep{Bahcalletal2005, AntiaBasu2006}.

To settle the open issues, a thorough exploration of the effects on
line formation of the interaction between magnetic field and
convection is a necessary step. In this letter we have presented a
study of that kind using three representative Fe lines. We have
determined that magnetic fields have a noticeable influence on the
formation of iron lines and consequently on the solar Fe abundance
determination.
For average vertical flux densities of $200$ G, we obtain abundance
corrections (to be applied in order to obtain the
same equivalent widths in HD and MHD) between $0.06$ and $0.09$ dex in
absolute value, while $|\Delta\log\epsilon_{\odot}(Fe)| \sim 0.05$ dex for all lines 
for the $100$ G case.
A flux density of order $10^2$ G, as in the range shown in Fig.~\ref{fig:ew},
is probably the relevant reference value since it has been shown to
correspond to the order of the mean field strength of the photospheric plasma 
\citep{TrujilloBueno+etal2004}.

Our results show that the direct (Zeeman) and indirect (stratification-related) effects
of the magnetic field contribute in different amounts to the changes in the different lines. The
indirect effects cause most of the equivalent width change of the visible Fe lines, with a
small contribution from the Zeeman effect; 
for the infrared line, in contrast, the Zeeman
effect is dominant, and in fact changes the sign of the abundance
correction. The absolute value of the total correction is similar for all
lines, so we can take it as the reference abundance change related with the
presence or absence of magnetic field in the convection model. In fact, 
for the visible lines, the use of  MHD models to match the observational data
will lead to {\it higher} derived iron abundance than when using an HD model,
since the predicted theoretical equivalent widths are smaller for the former
cases than for the latter.
The opposite applies for the \FeI\ $1564.85$ nm, since the increased
temperature of the MHD models is more than compensated by the very large
direct Zeeman line splitting.

There are different avenues to explore with a view to confirming and extending all these results.
One should consider further reliable Fe I line pairs
with different magnetic sensitivity \citep{VasShchu2010}, and carry
out an extensive study including all Stokes parameters for the same spectral features as in
previous 3D investigations \citep{ShchuTrujilloBueno2001, Aspetal2009}, but now based on MHD simulations.

On the other hand, spectral synthesis using 3D convection models should aim to match key
observational constraints such as spatially resolved line profiles
\citep{PerKisAsp2009,PerAspKis2009b}, center-to-limb continuum
behaviour, calibrated continuum intensities at disk center, spectral
energy distribution, and H lines. Finally, it is urgent to explore the 
magnetic effects in the Sun for lines of other crucial elements.

\acknowledgments

  Financial support by the European Commission through the SOLAIRE Network
  (MTRN-CT-2006-035484) and by the Spanish Ministry of Research and
  Innovation through projects AYA2007-66502, CSD2007-00050 and AYA2007-63881 is gratefully
  acknowledged, as are the computer resources, technical expertise and
  assistance provided by the MareNostrum (BSC/CNS, Spain), the Danish
  Center for Scientific Computing (DCSC/KU,
  Denmark), LaPalma (IAC/RES, Spain) and DEISA/HRLS (Germany) supercomputer installations.
  The work of {\AA}N was supported by the Danish Natural Science Research Council.
  We are grateful to A.~de~Vicente (IAC Condor management), H.~Socas-Navarro,
  C.~Allende-Prieto, B.~Ruiz~Cobo, N.~Shchukina, H.-G. Ludwig, Remo Collet, M. Asplund, J.~Trujillo Bueno, and
  J.~S\'anchez Almeida for their help and for sharing interesting discussions.


\begin{thebibliography}{}
\bibitem[Antia \& Basu(2006)]{AntiaBasu2006} Antia, H.~M., \& Basu, S. 2006,
    \apj, 644, 1292
\bibitem[Asplund et al.(2000a)]{Aspetal2000a} Asplund, M., Nordlund,
    \AA., Trampedach, R., Allende Prieto, C., \& Stein, R. F. 2000a, A\&A,
    359, 729
\bibitem[Asplund et al.(2000b)]{Aspetal2000b} Asplund, M., Ludwig, H.-G.,
    Nordlund, \AA., Stein, R.~F. 2000b, A\&A, 359, 669
\bibitem[Asplund(2005)]{Asp2005} Asplund, M. 2005, \araa, 43, 481
\bibitem[Asplund et al.(2009)]{Aspetal2009} Asplund, M., Grevesse, N.,
    Sauval, A.~J., \& Scott, P. 2009, ARA\&A, 47, 481
\bibitem[Ayres, Plymate, \& Keller(2006)]{AyrPlyKell(2006)} Ayres, T.~R., Plymate, C., \& Keller, C.~U. 2006, ApJ, 165, 618
\bibitem[Ayres(2008a)]{Ayres2008a} Ayres, T.~R. 2008, ASPC, 384, 52
\bibitem[Ayres(2008b)]{Ayres2008b} Ayres, T.~R. 2008, ApJ, 686, 731
\bibitem[Bahcall et al.(2005)]{Bahcalletal2005} Bahcall, J.~N., Basu, S.,
    Pinsonneault, M., \& Serenelli, A.~M. 2005, ApJ, 618, 1049
\bibitem[Blackwell, Lynas-Gray, \& Smith(1995)]{BlackLynSmith1995} Blackwell,
    D.~E., Lynas-Gray, A.~E., \& Smith, G. 1995, A\&A, 296, 217
\bibitem[Blackwell, Smith, \& Lynas-Gray(1995)]{BlackSmithLyn1995} Blackwell,
    D.~E., Smith, G., \& Lynas-Gray, A.~E. 1995, A\&A, 303, 575
\bibitem[Borrero(2008)]{Borrero2008} Borrero, J.~M. 2008, ApJ, 673, 470
\bibitem[Brault et al.(1987)]{Braultetal1987}Brault, J., \& Neckel, H. 1987, Spectral Atlas of the Solar Absolute Disk-averaged and Disk Center Intensity from 3290 \AA\ to 12510 \AA\ (Hamburg: Univ. Hamburg), ftp://nsokp.nso.edu/pub/atlas/
\bibitem[Caffau et al.(2008)]{Caffetal2008} Caffau, E., Ludwig, H.-G.,
    Steffen, M., Ayres, T.~R., Bonifacio, P., Cayrel, R., Freytag, B.,
    \& Plez, B. 2008, A\&A, 488, 1031
\bibitem[Caffau et al.(2010)]{Caffetal2010} Caffau, E., Ludwig, H.-G.,
    Steffen, M., Freytag, B., \& Bonifacio, P. 2010, Solar Phys. {\it [arXiv1003.1190]}
\bibitem[Carlsson et al.(2004)]{Carlssonetal2004} Carlsson, M., Stein, R.~F., Nordlund, \AA., \& Scharmer, G.~B. 2004, ApJ, 610, L137
\bibitem[De Pontieu et al.(2006)]{DePontieuetal2006} De Pontieu, B., Carlsson, M., Stein, R., Rouppe van der Voort, L., L\"{o}fdahl, M., van Noort, M., Nordlund, \AA., \& Scharmer, G. 2006, ApJ, 646, 1405
\bibitem[Dravins et al.(1981)]{Dravinsetal1981} Dravins, D.,
    Lindegren, L., Nordlund, \AA. 1981, A\&A, 96, 345
\bibitem[Fabbian et al.(2006)]{Fabbianetal2006} Fabbian, D., Asplund, M.,
    Carlsson, M., \& Kiselman, D. 2006, A\&A, 458, 899
\bibitem[Fabbian et al.(2009)]{Fabbianetal2009} Fabbian, D., Asplund, M.,
    Barklem, P.~S., Carlsson, M., \& Kiselman, D. 2009, A\&A, 500, 1221
\bibitem[Freytag et al.(2002)]{Freytagetal2002} Freytag, B., Steffen, M.,
    \& Dorch, B. 2002, Astron. Nachr., 323, 213
\bibitem[Gray(1992)]{Gray1992} Gray, D.~F. 1992, "The observation and analysis of stellar photospheres" (second edition), Camb. Astrophys. Ser., Vol. 20
\bibitem[Gurtovenko \& Kostik(1986)]{GurtovenkoKostik} Gurtovenko, E. A. \& Kostik, R. I. 1986, Naukova Dumka, Kiev
\bibitem[Holweger, Kock, \& Bard(1995)]{HolwKockBard1995} Holweger, H.,
    Kock, M., \& Bard, A. 1995, A\&A, 296, 233
\bibitem[Khomenko et al.(2003)]{Khomenko+etal2003} Khomenko, E., Collados, M., Solanki, S. K., Lagg, A. \& Trujillo Bueno, J.
 2003, A\&A, 408, 1115
\bibitem[Kostik, Shchukina \& Rutten(1996)]{KostShchuRutt1996} Kostik,
    R.~I., Shchukina, N.~G., \& Rutten, R.~J., 1996, A\&A, 305, 325
\bibitem[Mart\'inez Gonz\'alez \& Bellot Rubio(2009)]{MarGonBellRub2009} Mart\'inez Gonz\'alez, M.~J. \& Bellot Rubio L.~R. 2009, ApJ, 700, 1391
\bibitem[Nordlund(1985)]{Nordlund1985} Nordlund, \AA., 1985, Solar Phys., 100, 209
\bibitem[{{Nordlund} \& {Galsgaard}(1995)}]{nordlund+galsgaard95}
{Nordlund}, A. \& {Galsgaard}, K. 1995, {Tech. Rep., Astron. Observ.,
  Copenhagen Univ., http://www.astro.ku.dk/~aake/papers/95.ps.gz}
\bibitem[Nordlund, Stein, \& Asplund(2009)]{NordStAsp2009} Nordlund, \AA.,
    Stein, R.~F.; Asplund, M. 2009, LRSP, 6, 2
\bibitem[Pereira, Kiselman \& Asplund (2009)]{PerKisAsp2009} Pereira, T.~M.~D., Kiselman, D.,
    \& Asplund, M. 2009, A\&A, 507, 417
\bibitem[Pereira, Asplund \& Kiselman (2009a)]{PerAspKis2009a}
   Pereira, T.~M.~D., Asplund, M., \& Kiselman, D. 2009, IAU General
   Assembly Joint Discussion 10 "3D views on cool stellar atmospheres:
   theory meets observation", {\it (to appear in MmSAI), arXiv0909.4121}
\bibitem[Pereira, Asplund \& Kiselman (2009b)]{PerAspKis2009b} Pereira, T.~M.~D., Asplund, M.,
    \& Kiselman, D. 2009, A\&A, 508, 1403
\bibitem[S\'anchez Almeida, Emonet \& Cattaneo(2003)]{SanchezAlmeida_etal_2003} S\'anchez Almeida, J., Emonet, T., \& Cattaneo, F. 2003, ApJ, 585, 536
\bibitem[S\'anchez Almeida et al.(2004)]{SanchezAlmeida_etal_2004}
   S{\'a}nchez Almeida, J., M{\'a}rquez, I., {Bonet}, J.~A.,
Dom{\'{\i}}nguez Cerde{\~n}a, I., {Muller}, R., 2004, ApJ, 609, L91
\bibitem[S\'anchez Almeida et al.(2010)]{SanchezAlmeida_etal_2010}
   S{\'a}nchez Almeida, J., {Bonet}, J.~A., Viticchi\'e, B., Del Moro, D., 2010, A\&A, 715, L26
\bibitem[Shchukina \& Trujillo Bueno(2001)]{ShchuTrujilloBueno2001} Shchukina, N.,
    \& Trujillo Bueno, J. 2001, ApJ, 550, 970
\bibitem[Shchukina \& Trujillo Bueno(2003)]{Shchukina_TrujilloBueno2003} Shchukina, N., \& Trujillo Bueno, J. 2003, ASPC, 307, 336
\bibitem[Shelyag et al. (2004)]{shelyag_etal_2004} Shelyag, S., Sch\"ussler,
  M., Solanki, S.~K., Berdyugina, S.~V., V\"ogler, A. 2004, ApJ 427, 335.
\bibitem[Socas-Navarro(2001)]{Socas-Navarro2001} Socas-Navarro, H. 2001, ASPC, 236, 487
%
%
\bibitem[Socas-Navarro \& Norton(2007)]{Socas-Navarro&Norton2007} Socas-Navarro, H., \& Norton, A.~A. 2007, ApJ, 660, L153
\bibitem[Sch\"uessler \& Solanki(1988)]{SchSol1988} Sch\"uessler, M. \& Solanki, S.~K. 1988, A\&A, 192, 338
\bibitem[Spruit(1976)]{Spruit1976} Spruit, H.~C. 1976, Solar Phys., 50, 269
\bibitem[Stein \& Nordlund(2006)]{Stein&Nord2006} Stein, R.~F., \& Nordlund, \AA. 2006, ApJ, 642, 1246
\bibitem[Stenflo \& Lindegren(1977)]{Stenflo1977} Stenflo, J. O \& Lindegren, L. 1977, A\&AS, 59,
367
\bibitem[Stenflo(1982)]{Stenflo1982} Stenflo, J.~O. 1982, Solar Phys., 80, 209
\bibitem[Trujillo Bueno, Shchukina \& Asensio Ramos(2004)]{TrujilloBueno+etal2004} Trujillo Bueno, J., Shchukina, N.,
    \& Asensio Ramos, A. 2004, Nature, 430, 326
\bibitem[Trujillo Bueno \& Shchukina(2009)]{TruShchu2009} Trujillo
    Bueno, J. \& Shchukina, N.~G., 2009, \apj, 694, 1364
\bibitem[Vasilyeva \& Shchukina(2010)]{VasShchu2010} Vasilyeva, I. E., \&
    Shchukina, N. G. 2010, KPCB, 25, 319
\bibitem[V\"{o}gler et al.(2005)]{Vogleretal2005} V\"{o}gler, A., Shelyag, S.,
    Sch\"{u}ssler, M., Cattaneo, F., Emonet, T., \& Linde, T. 2005, A\&A, 429, 335

\end{thebibliography}
\end{document}